\documentclass[twocolumn,twoside,preprintnumbers,amsmath,amssymb,pacs]{revtex4}
\usepackage{epsfig}
\usepackage{graphicx}
\usepackage{dcolumn}
\usepackage{bm}
\usepackage{fancyhdr}
\usepackage{pslatex}

\begin{document}

\title{{\Large  Gauge/string duality  
 and hadronic physics   }}

\author{Henrique Boschi-Filho and Nelson R. F. Braga}

\affiliation{Instituto de F\'{\i}sica, 
Universidade Federal do Rio de Janeiro, Caixa Postal 68528, 
RJ 21941-972 -- Brazil}

\received{on 24 March, 2006}

\begin{abstract}
We review some recent results on phenomenological approaches to strong interactions inspired in gauge/string duality. In particular, we discuss how such models lead  to very good estimates for hadronic masses.  


\end{abstract}

\maketitle

\thispagestyle{fancy}
\setcounter{page}{0}

\section{Introduction}

Let us start by briefly reviewing some early results on strong interactions
which lead to the proposal of string theory.
These results date back to the decade of 1960.
The first is the experimental observation, in hadronic scattering, of an apparently infinite tower of resonances with mass and angular momenta related by 

\begin{equation}
\label{1}
J \,\,\sim\,\,m^2 \,\alpha^\prime 
\end{equation}

\noindent where $\alpha^\prime \sim 1 (GeV)^{-2} $ is the Regge slope.  

Another important fact is that the properties of hadronic scattering 
in the so called Regge regime are nicely described by the amplitude 
postulated by Veneziano (in terms of Mandelstan variables ($s,t,u$))

\begin{equation}
\label{2}
 A(s,t) \,=\,{ \Gamma (-\alpha (s) )\, \Gamma (-\alpha (t) ) \over
 \Gamma (-\alpha (s) - \alpha (t) ) }
\end{equation}

\noindent where $ \Gamma (z)\,$ is the Euler gamma function and 
$ \alpha (s) \,=\,\alpha(0) + \alpha^\prime s $.

A strong motivation for relating hadrons to strings is the
fact that these two results of eqs. (\ref{1}) and (\ref{2}) can be reproduced 
from a relativistic bosonic string\cite{GSW,Polchinski:1998rq}.    
A relativistic bosonic string can be described by the action

\begin{equation}
S\,=\, -{1\over 4\pi \alpha^\prime}
\int d\tau d\sigma \sqrt{ -g} g^{ab} \partial_a X^\mu \partial_b X_\mu \,.
\end{equation}

\noindent After quantization, one finds that the spectrum of 
excitations shows up in representations satisfying eq. (\ref{1}). 
The string scattering amplitudes also reproduce the Veneziano result eq. (\ref{2}).

On the other hand, some problems concerning a possible string description
of hadronic physics have been a challenge for physicists for a long time.
One of them is the behavior of the scattering amplitudes at high energy.
If one considers the Regge limit, corresponding to $ s \rightarrow \infty \,,\,$ 
with fixed  $\, t \,:$ the Veneziano amplitude behaves as 

\begin{equation}
 A \sim s^{\alpha(t)}\,, 
\end{equation}

\noindent in agreement with experimental results. Actually this was one 
of the inputs for building up the Veneziano amplitude.
However considering high energy scattering at fixed angles that correspond to the
limit $ s \rightarrow \infty \,\,\, $ with $\,\,s/t\,$ fixed the amplitude behaves as
(soft scattering)

\begin{equation}
\,\,\, A_{_{Ven.}} \sim exp\,\{\, -\alpha^\prime s f (\theta)\, \} 
\end{equation}

\noindent while experimental results for hadrons show  a hard scattering behavior

\begin{equation}
\label{6}
 A_{exp.} \sim \, s^{(4-\Delta )/2} \,\,,
\end{equation}

\noindent that is reproduced by QCD\cite{QCD1,BRO}. 

As we will discuss later, this apparent obstacle to describing strong interactions using string theory was solved recently by Polchinski and Strassler\cite{PS}.

In the mean time it was realized that string theory contains massless
spin two excitations that can be associated with the graviton.
This allows one to interpret string theory as a quantum theory for gravity.
In fact the symmetry groups for the possible string theories are large enough
to contain also all the fields of the standard model. So presently string theory 
is a candidate for a unified theory for all interactions. 

In this review we are going to discuss recent results concerning the relation between 
string theory and QCD. QCD has been tested and confirmed with success in high energy 
experiments. But QCD is non perturbative at low energies.
Lattice calculations give us very important results in this regime.
However it seems that we are still far from a complete description of the 
complexity of strong interactions. In particular, important aspects 
like confinement and mass generation still lack of a satisfactory description.
Presently there are many indications that string theory can be useful
in the description of strong interactions in the non perturbative regimes  
of QCD.
 
An early connection between SU(N) gauge theories (for large N) and string theory
was realized long ago by 't Hooft\cite{'tHooft:1973jz}.
A few years ago a very important result was obtained by Maldacena\cite{Malda}
relating string theory in anti-de Sitter (AdS) space with gauge theories in Minkowski
space. 
He established a correspondence between string theory in $AdS_5 \times S^5 \,$ 
space-time and ${\cal N } = 4$ superconformal Yang Mills SU(N) theory for large N 
in its four dimensional boundary. This is known as AdS/CFT 
correspondence\cite{Malda,GKP,Wi}. 

In the AdS/CFT correspondence we have an exact duality between a four dimensional gauge theory and string theory in a ten dimensional space.
However, in this formulation, the gauge theory has no energy scale as it is conformal.  
There are many attempts to search for exact dualities involving gauge theories more
similar to QCD\cite{Maldacena:2000yy,Klebanov:2000nc,Klebanov:2000hb,Polchinski:2000uf}.

Although it involves a conformal gauge theory, the AdS/CFT correspondence has been a very important source of inspiration for searching QCD results from string theory.
The first idea of breaking conformal invariance in the AdS/CFT context,
proposed by Witten, is to consider an AdS Schwarzschild black hole as dual to a non-supersymmetric Yang Mills theory\cite{Wi2}. This approach was used to
calculate glueball masses in \cite{MASSG,MASSG2,MASSG3,MASSG4,MASSG5,MASSG6,MASSG7}.

\section{QCD scattering and string theory}

Recently Polchinski and Strassler\cite{PS} introduced an infrared cut off in 
the AdS space and reproduced the hard scattering behavior of strong interactions at fixed angles from string theory.
Inspired in the AdS/CFT correspondence they assumed 
a duality between gauge theory glueballs and string theory dilatons in an AdS slice
and found the QCD scaling.

This scaling was also obtained in \cite{BB3} from a  mapping between quantum states 
in AdS space and its boundary found in \cite{BB2}. 
We considered an AdS slice as approximately dual to a confining gauge 
theory.  The slice corresponds to the metric

\begin{equation}
ds^2=\frac {R^2 }{( z )^2}\Big( dz^2 \,+(d\vec x)^2\,
- dt^2 \,\Big)\,,
\end{equation}

\noindent with $\, 0\le z \le  z_{max} \,\, \sim 1/\mu $  where  $\mu$
is an energy scale chosen as the mass of the lightest glueball.
We used a mapping between Fock spaces of a scalar field in AdS space and operators 
on the four dimensional boundary, defined in \cite{BB2}. 
Considering a scattering of two particles in $m$ particles 
one finds a relation between bulk and boundary scattering amplitudes\cite{BB3}

\begin{equation}
S_{Bulk} \, \sim \,  
 S_{Bound.} \,\,\Big( {\sqrt{\alpha^\prime} \over \mu }\Big)^{m+2} \,\, K^{(m+2)(1+d)}
\end{equation}

\noindent where $d $ is the scaling dimension of the boundary operators and $K$ is the boundary momentum scale. This leads to the result for the amplitude

\begin{equation}
A_{Boundary} \,\sim \,s^{(4  - \Delta)/2 } \,,
\end{equation}

\noindent where $\Delta$ is the total scaling dimension 
of scattered particles. This reproduces the QCD scaling of eq. (\ref{6}).

For some other results concerning QCD scattering properties from string theory see also 
\cite{GI,BT,AN,PS2,Brodsky:2003px,AN2,AN3}. 
 
\section{Scalar glueball masses }

Using the phenomenological approach of introducing an energy scale 
by considering an AdS slice we found estimates for scalar glueball mass ratios\cite{BB4,BB5}.
In the AdS$_5$ bulk we took dilaton fields satisfying Dirichlet boundary conditions at $ z = z_{max}$ 
\begin{eqnarray}
\label{QF}
\Phi(z,\vec x,t) &=& \sum_{p=1}^\infty \,
\int { d^3 k \over (2\pi)^{3}}\,
{z^{2} \,J_2 (u_p z ) \over z_{max}\,\, w_p(\vec k ) 
\,J_{3} (u_p z_{max} ) }\nonumber\\
&\times& \lbrace { {\bf a}_p(\vec k )\ }
 e^{-iw_p(\vec k ) t +i\vec k \cdot \vec x}\,
\,+\,\,h.c.\rbrace\,,\nonumber
\end{eqnarray}

\noindent where $\,w_p(\vec k ) \,=\,
\sqrt{ u_p^2\,+\,{\vec k}^2}\,$, 

\begin{equation}
 u_p \,=\,\frac{ \chi_{_{2\,,\,p}}}{z_{max}} \,,
\end{equation}

\noindent is the momentum associated with the $z$ direction
and $\chi_{_{2\,,\,p}} $ are the zeroes of the Bessel functions:  
$\,J_2 (\chi_{_{2\,,\,p}} )=0\,.$
 
 On the boundary ($ z = 0)$ we considered scalar glueball states  $J^{PC}\,=\,0^{++}$ 
and their excitations $0^{++\ast},\,\,0^{++\ast\ast}$  with masses 
$\mu_p\,,\,p=1,2,...$. 
Assuming an approximate gauge/string duality the 
glueball masses are taken as proportional to the dilaton discrete modes: 
$$
{u_p \over \mu_p }\,=\,{\rm const.}\,
$$

\noindent So, the ratios of glueball masses are related  
to zeros of the Bessel functions
$$
{ \mu_p\over \mu_1 }\,=\,{\chi_{2\,,\,p}\over \chi_{2\,,\,1}}\,\,.
$$

\bigskip

\noindent Note that these ratios are independent of the size of the slice   $\,\,\,\,\,z_{max}\,$. 
Our estimates compared with $SU(3) $ lattice \cite{LAT1,LAT2} and AdS-Schwarzshild 
\cite{MASSG} (in GeV) are shown in Table I.

\begin{table}[htbp]
\begin{tabular}{|l|c|c|c|}
\hline
4d State  & {  lattice}, $N=3$ &
{  AdS-BH} & {  AdS slice} \\
 \hline
 $0^{++}$ & $1.61 \pm 0.15$   & 1.61  & 1.61  \\
 $0^{++*}$ &  2.8   & 2.38 & 2.64 \\
 $0^{++**}$ &   - & 3.11 & 3.64 \\
 $0^{++***}$ &  -  & 3.82 & 4.64\\
 $0^{++****}$ &  -  & 4.52 & 5.63\\
 $0^{++*****}$ &  -  & 5.21 & 6.62\\
\hline
\end{tabular}
\caption{Four dimensional glueball masses in GeV with Dirichlet boundary conditions. The value 1.61 of the third and fourth columns is an input taken from lattice results.}
\end{table}

\begin{table}[htbp]
\begin{tabular}{|l|c|c|c|c|}
\hline
3d State & {  lattice, $N=3$} & 
{  lattice,} {  $ N \rightarrow \infty $ } &
{  AdS-BH} & {  AdS slice} \\
 \hline
 $0^{++}$ & $4.329 \pm 0.041$ & $4.065 \pm 0.055$ & 4.07 
& 4.07  \\
 $0^{++*}$ & $6.52 \pm 0.09$ & $6.18 \pm 0.13$ & 7.02 & 7.00\\
 $0^{++**}$ & $8.23 \pm 0.17$ & $7.99 \pm 0.22$ & 9.92 & 9.88 \\
 $0^{++***}$ &  - & - & 12.80 & 12.74 \\
 $0^{++****}$ &  - & - & 15.67 & 15.60\\
 $0^{++*****}$ & -  & - & 18.54 & 18.45\\
\hline
\end{tabular}
\caption{Three dimensional glueball masses in units of string tension with Dirichlet boundary conditions. The value 4.07
is an input from lattice.}
\end{table}

A similar approach was used also for glueball masses in QCD$_3$, taken as dual to 
scalar fields is AdS$_{4}$. In this case the Bessel functions are $J_{3/2} $ 
and  the mass ratios take the form 
\begin{equation}
{ \mu_p\over \mu_1 }\,=\,{\chi_{3/2\,,\,p}\over \chi_{3/2\,,\,1}}\,\,.
\end{equation}
 
\noindent Our results for QCD$_3$ are shown in Table II, again 
compared with lattice \cite{LAT1,LAT2} and AdS-Schwarzshild 
\cite{MASSG} results.
 
For some other results concerning glueball masses using gauge/string duality
see for instance \cite{Caceres:2000qe,ACEP,Amador:2004pz,Caceres:2005yx}.

\section{ Higher spin states and Regge trajectories}

Recently, very interesting results for the hadronic
spectrum were obtained by Teramond and Brodsky\cite{deTeramond:2005su} considering scalar, vector and fermionic fields in a sliced $ AdS_5 \times S^5 $ space.
It was proposed that massive  bulk states corresponding to fluctuations about 
the $AdS_5$ metric are dual to QCD states with  
angular momenta (spin) on the four dimensional boundary.
This way the spectrum of light baryons and mesons has been reproduced
from a holographic dual to QCD inspired in the AdS/CFT 
correspondence.

We used a similar approach to estimate masses of glueball states 
with different spins\cite{Boschi-Filho:2005yh}. 
The motivation was to compare the glueball Regge trajectories with
the pomeron trajectories. For soft pomerons \cite{Landshoff:2001pp} experimental
results show that
\begin{equation}
\label{11}
J \,\approx \,  1.08 \,+\, 0.25\, M^2\,\,\,\,\,\,\,\,\,(GeV)\,.
\end{equation}

\noindent It is conjectured that the soft pomerons may be related to glueballs.
Recent lattice results are consistent with this interpretation\cite{Meyer:2004jc}.
 
We assume that massive scalars in the AdS slice with mass $\mu$ 
are dual to boundary gauge theory states with spin $J$ related by: 
\begin{equation}
( \mu R )^2 \,=\, J ( J + 4 ) \,\,.
\end{equation}

We consider both Dirichlet and Neumann boundary conditions and the results for the
four dimensional glueball masses with even spin are shown in tables III and IV respectively.

\begin{table}
\begin{tabular}{ | c | c | c | c |} 
\hline 
Dirichlet $\,\,$ 
 & $\,\,$ lightest $\,\,$   &
$1^{st}$ excited $\,\,$ & $2^{nd}$ excited \\
glueballs $\,\,$     &   state & state & state \\
 \hline
 $0^{++}$ & 1.63  &  2.67 & 3.69 \\ 
 $2^{++}$ & 2.41    & 3.51  & 4.56  \\
 $4^{++}$ & 3.15  & 4.31  & 5.40  \\
 $6^{++}$ & 3.88  & 5.85  & 6.21 \\
 $8^{++}$ & 4.59 & 5.85  & 7.00 \\
 $10^{++}$ & 5.30 & 6.60   & 7.77 \\
\hline
\end{tabular}
\caption{     Higher spin glueball masses in GeV with Dirichlet boundary condition. 
The value 1.63 is an input from lattice.}
\end{table}

\begin{table}
\begin{tabular}{ | c | c | c | c |} 
\hline 
Neumann $\,\,$ 
 & $\,\,$ lightest $\,\,$   &
$1^{st}$ excited $\,\,$ & $2^{nd}$ excited \\
glueballs $\,\,$     &   state & state & state \\
 \hline
 $0^{++}$ & 1.63  & 2.98  & 4.33  \\ 
 $2^{++}$ & 2.54    & 4.06 & 5.47  \\
 $4^{++}$ & 3.45 & 5.09  & 6.56 \\
 $6^{++}$ & 4.34  & 6.09 & 7.62 \\
 $8^{++}$ & 5.23 & 7.08 & 8.66 \\
 $10^{++}$ & 6.12 & 8.05  & 9.68 \\
\hline
\end{tabular}
\caption{  Higher spin glueball masses in GeV with Neumann boundary condition. 
The value 1.63 is an input from lattice.  }
\end{table} 
   
We found non linear relations between spin and mass squared.
We considered linear approximations representing Regge trajectories
\begin{equation}
 J \,=\, \alpha_0 \,+\,\alpha^{\prime} \, M^2\,.
\end{equation}

\noindent For Neumann boundary conditions for the states  
$J^{++}\,$ with $J = 2,4,...,10\,$ we found 
\begin{equation}
\alpha^{\prime}\,=\,(\,0.26 \pm 0.02 \,)GeV^{-2} 
\qquad ; \qquad \alpha_0 \,=\,0.80 \pm 0.40\,,
\end{equation}

\noindent as shown in Figure 1.

\begin{figure}[htbp]
\begin{center}
\includegraphics[width=8cm]{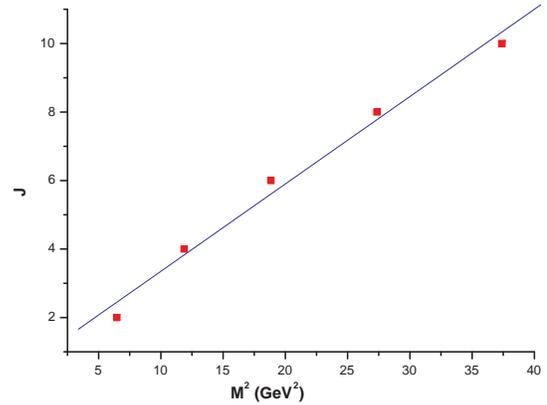}
\caption{  Spin versus mass squared for the lightest glueball states with 
Neumann boundary conditions from table IV. The line corresponds to the linear fit.}
\end{center}
\end{figure}

\begin{figure}[htbp]
\begin{center}
\includegraphics[width=8cm]{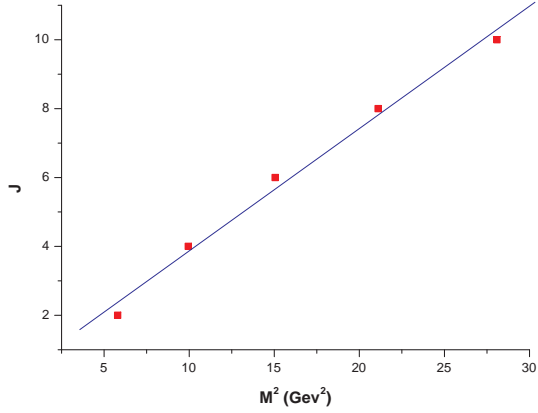}
\caption{ Spin versus mass squared for the lightest glueball states with 
 Dirichlet boundary conditions from table III. 
The line corresponds to the linear fit. }
\end{center}
\end{figure}
\noindent For Dirichlet boundary conditions, taking the states 
$J^{++}\,$ with $J = 2,4,...,10\,$ 
we found a linear fit with 
\begin{equation}
\alpha^{\prime}\,=\,( \, 0.36 \pm  \,0.02\,)\,GeV^{-2}\, \qquad ; \qquad
 \alpha_0 \,=\,0.32 \pm 0.36 \,,
\end{equation}

\noindent as shown in Figure 2.

So, Neumann boundary conditions give a glueball trajectory consistent with
that of pomerons, eq. (\ref{11}). These kind of boundary conditions appear in the Randall Sundrum 
model\cite{Randall:1999ee} as a consequence of the orbifold condition.
    
\section{Wilson loops and quark anti-quark potential}

Confinement criteria for QCD can be discussed with the help of Wilson loops
which give the binding energy of the system.
Phenomenological results imply that the quark anti-quark potential  
has the form

\begin{equation}
\label{Cornell}
E_{Cornell}(L) \,=\, -\frac{4}{3} \frac{a}{L} \,+\, \sigma L\,+\, constant\,\,,
\end{equation}
\noindent where , $a = 0.39\,$ and $\sigma = 0.182 GeV^2\,$. 

In the AdS/CFT correspondence Wilson loops for a heavy quark anti-quark pair in the
conformal gauge theory can be calculated from a static string in the 
AdS space\cite{RY,MaldaPRL}. The corresponding energy is a non confining Coulomb
potential. For an excellent review and extension to other metrics
see \cite{Kinar:1998vq}.  

We calculated Wilson loops for a quark anti-quark pair in D3-brane space finding different confining behaviors depending on the quark 
position\cite{Boschi-Filho:2004ci}.

Recently we calculated\cite{Boschi-Filho:2005mw} the energy of a static string in an
AdS slice between two D3-branes with orbifold condition. 
The energy for configurations with endpoints
on a brane grows linearly for large separation between these points. 
The derivative of the energy has a discontinuity at some critical 
separation.  Choosing a particular position for one of the branes 
we find configurations with smooth energy. 
In the limit where the other brane goes to infinity the energy 
has a Coulombian behaviour for short separations 
and can be identified with the Cornell potential eq. (\ref{Cornell}).
This identification leads to effective  
values for the string tension, the position of the infrared brane and the AdS radius
\begin{equation}
R \,=\,\sqrt{ \frac{a}{ 3 \sigma C_1^2}}\,=\, 1.4\,\, {\rm GeV}^{-1}\,\,,
\end{equation}
\noindent where $C_1 = \sqrt{2} \pi^{3/2}/[\Gamma(1/4)]^2\,$.

These results suggest an approximate duality 
between static strings in an AdS slice and a heavy quark anti-quark 
configuration in a confining gauge theory.

For other interesting results concerning gauge/string duality and QCD see for 
instance  \cite{Janik:1999zk,Janik:2000aj,Janik:2001sc, PandoZayas:2003yb,Andreev:2004sy,Bigazzi:2004ze,Erlich:2005qh,
DaRold:2005zs,Evans:2005ip,Brodsky:2005kc,Brodsky:2006uq}.

\acknowledgements{ The authors are partially supported by
CNPq and Faperj.}


\begin{thebibliography}{99}

\bibitem{GSW}
  M.~B.~Green, J.~H.~Schwarz and E.~Witten,
``Superstring Theory. Vol. {\bf 1}: Introduction,'', Cambridge, 1987.

\bibitem{Polchinski:1998rq}
  J.~Polchinski,
  ``String theory. Vol. {\bf 1}: An introduction to the bosonic string,'' Cambridge 1998.


\bibitem{QCD1} V. A. Matveev, R.M. Muradian and A. N. Tavkhelidze,
 Lett. Nuovo Cim. {\bf 7} (1973) 719.

\bibitem{BRO} S. J. Brodsky and G. R. Farrar, Phys. Rev. Lett 31 (1973) 1153;
Phys. Rev. {\bf D11} (1975) 1309.


\bibitem{PS} 
J. Polchinski and M. J. Strassler, Phys. Rev. Lett. {\bf  88} (2002) 031601.

\bibitem{'tHooft:1973jz}   G.~'t Hooft,
  Nucl.\ Phys.\ B {\bf 72} (1974)  461.

\bibitem{Malda} J. Maldacena, Adv. Theor. Math. Phys. {\bf 2} (1998) 231.

\bibitem{GKP} S. S. Gubser , I.R. Klebanov and A.M. Polyakov, 
Phys. Lett. {\bf B428} (1998) 105.

\bibitem{Wi} E. Witten, Adv. Theor. Math. Phys. 2 (1998) 253.

\bibitem{Maldacena:2000yy}
J.~M.~Maldacena and C.~Nunez,
Phys.\ Rev.\ Lett.\  {\bf 86} (2001)  588.

\bibitem{Klebanov:2000nc}
I.~R.~Klebanov and A.~A.~Tseytlin,
Nucl.\ Phys.\ B {\bf 578} (2000) 123.

\bibitem{Klebanov:2000hb}
I.~R.~Klebanov and M.~J.~Strassler,
JHEP {\bf 0008} (2000)  052.

\bibitem{Polchinski:2000uf}
J.~Polchinski and M.~J.~Strassler,
``The string dual of a confining four-dimensional gauge theory,''
arXiv:hep-th/0003136.


\bibitem{Wi2} E. Witten, Adv.Theor.Math.Phys. {\bf 2} (1998) 505.


\bibitem{MASSG} C. Csaki, H. Ooguri, Y. Oz and J. Terning, JHEP {\bf 9901} (1999) 017.

\bibitem{MASSG2} R. de Mello Koch, A. Jevicki, M. Mihailescu , J. P. Nunes,  
Phys.Rev. {\bf D58} (1998)105009.

\bibitem{MASSG3} A. Hashimoto , Y. Oz , Nucl.Phys. {\bf B548} (1999) 167. 

\bibitem{MASSG4} C. Csaki , Y. Oz , J. Russo , J. Terning , 
Phys.Rev. {\bf D59} (1999) 065012. 

\bibitem{MASSG5} J. A. Minahan,  JHEP {\bf 9901} (1999) 020.

\bibitem{MASSG6} C. Csaki, J. Terning,  AIP Conf. Proc. {\bf 494} (1999) 321.
  
\bibitem{MASSG7} R. C. Brower, S. D. Mathur , C. I. Tan , Nucl.Phys.B587(2000) 249. 

\bibitem{BB3} H. Boschi-Filho and N. R. F. Braga, Phys. Lett. {\bf B560} (2003)
232. 

\bibitem{BB2} H. Boschi-Filho and N. R. F. Braga, Phys. Lett. {\bf B525} (2002) 164.

\bibitem{GI} 
S. B. Giddings, 
Phys.\ Rev.\ {\bf D 67 } (2003) 126001. 

\bibitem{BT} 
R. C. Brower , C. I. Tan, 
Nucl.\ Phys.\ {\bf B 662} (2003) 393.

\bibitem{AN} 
  O.~Andreev,
  Phys.\ Rev.\ D {\bf 67} (2003) 046001.

\bibitem{PS2}
  J.~Polchinski and M.~J.~Strassler,
  JHEP {\bf 0305 } (2003) 012.

\bibitem{Brodsky:2003px}
  S.~J.~Brodsky and G.~F.~de Teramond,
  Phys.\ Lett.  {\bf B 582} (2004)  211.

\bibitem{AN2}
  O.~Andreev,
  Phys.\ Rev.\ {\bf D 70} (2004) 027901.


\bibitem{AN3}
  O.~Andreev,
  Phys.\ Rev.\ D {\bf 71} (2005)  066006.

\bibitem{BB4}
  H.~Boschi-Filho and N.~R.~F.~Braga,
  Eur.\ Phys.\ J.\ C {\bf 32} (2004) 529.


\bibitem{BB5} H. Boschi-Filho and N. R. F. Braga, JHEP {\bf 0305} (2003) 009.

\bibitem{LAT1} 
C. J. Morningstar and M. Peardon, Phys. Rev. D 56 (1997) 4043.

\bibitem{LAT2} M.J. Teper, "Physics from lattice: Glueballs in QCD; 
topology; SU(N) for all N ", arXiv:hep-lat/9711011.

\bibitem{Caceres:2000qe}
  E.~Caceres and R.~Hernandez,
  Phys.\ Lett.\ B {\bf 504} (2001)  64.

\bibitem{ACEP}  R. Apreda, D. E. Crooks, N. Evans, M. Petrini, 
JHEP {\bf 0405} (2004) 065.

\bibitem{Amador:2004pz}
  X.~Amador and E.~Caceres,
  JHEP {\bf 0411} (2004) 022.

\bibitem{Caceres:2005yx}
  E.~Caceres and C.~Nunez,
JHEP {\bf 0509} (2005)  027. 

\bibitem{deTeramond:2005su}   G.~F.~de Teramond and S.~J.~Brodsky,
Phys. Rev. Lett. {\bf 94} (2005) 201601.

\bibitem{Boschi-Filho:2005yh}
  H.~Boschi-Filho, N.~R.~F.~Braga and H.~L.~Carrion,
  Phys.\ Rev.\ D {\bf 73} (2006) 047901.

\bibitem{Landshoff:2001pp}
  P.~V.~Landshoff,
  ``Pomerons,'', published in  ``Elastic and Difractive Scattering" 
 Proceedings, Ed. V. Kundrat and P. Zavada, 2002, arXiv:hep-ph/0108156.

\bibitem{Meyer:2004jc}
  H.~B.~Meyer and M.~J.~Teper,
  Phys.\ Lett.\ B {\bf 605} (2005) 344.

\bibitem{Randall:1999ee}
  L.~Randall and R.~Sundrum,
  Phys.\ Rev.\ Lett.\  {\bf 83}, 3370 (1999);
ibid.  {\bf 83}, 4690 (1999).

\bibitem{RY} S.~J.~Rey and J.~T.~Yee,
Eur.\ Phys.\ J.\ C {\bf 22} (2001) 379.
  
\bibitem{MaldaPRL} J.~Maldacena, 
Phys.\ Rev.\ Lett.\  {\bf 80} (1998)  4859.

\bibitem{Kinar:1998vq}
Y.~Kinar, E.~Schreiber and J.~Sonnenschein,
Nucl.\ Phys.\ B {\bf 566} (2000) 103.

\bibitem{Boschi-Filho:2004ci}
  H.~Boschi-Filho and N.~R.~F.~Braga,
  JHEP {\bf 0503} (2005)  051.

\bibitem{Boschi-Filho:2005mw}
  H.~Boschi-Filho, N.~R.~F.~Braga and C.~N.~Ferreira,
  arXiv:hep-th/0512295.


\bibitem{Janik:1999zk}
  R.~A.~Janik and R.~Peschanski,
  Nucl.\ Phys.\ B {\bf 565} (2000) 193.

\bibitem{Janik:2000aj}
  R.~A.~Janik and R.~Peschanski,
  Nucl.\ Phys.\ B {\bf 586} (2000) 163.

\bibitem{Janik:2001sc}
  R.~A.~Janik and R.~Peschanski,
  Nucl.\ Phys.\ B {\bf 625} (2002) 279.


\bibitem{PandoZayas:2003yb}
  L.~A.~Pando Zayas, J.~Sonnenschein and D.~Vaman,
  Nucl.\ Phys.\ B {\bf 682} (2004) 3.


\bibitem{Andreev:2004sy}
  O.~Andreev and W.~Siegel,
  Phys.\ Rev.\ D {\bf 71} (2005)  086001.

\bibitem{Bigazzi:2004ze}
  F.~Bigazzi, A.~L.~Cotrone, L.~Martucci and L.~A.~Pando Zayas,
  Phys.\ Rev.\ D {\bf 71} (2005) 066002.

\bibitem{Erlich:2005qh}
  J.~Erlich, E.~Katz, D.~T.~Son and M.~A.~Stephanov,
Phys. Rev. Lett. {\bf 95} (2005) 261602.

\bibitem{DaRold:2005zs}
  L.~Da Rold and A.~Pomarol,
 Nucl.\ Phys.\  {\bf B 721} (2005) 79.


\bibitem{Evans:2005ip}
  N.~Evans, J.~P.~Shock and T.~Waterson,
Phys.Lett. {\bf B622} (2005) 165.

\bibitem{Brodsky:2005kc}
  S.~J.~Brodsky and G.~F.~de Teramond,
  AIP Conf.\ Proc.\  {\bf 814} (2006) 108.
  [arXiv:hep-ph/0510240].

\bibitem{Brodsky:2006uq}
  S.~J.~Brodsky and G.~F.~de Teramond,
  arXiv:hep-ph/0602252.






\end{thebibliography}
\end{document}